\documentclass[11pt]{article}
\usepackage{graphicx}
\usepackage{amssymb}
\usepackage{psfig}

\title{MAVID: Constrained ancestral alignment of multiple sequences}
\author{Nicolas Bray\thanks{Department of Mathematics, U.C. Berkeley} \and Lior Pachter\footnotemark[1]}
\begin{document}
\maketitle
\begin{abstract}
We describe a new global multiple alignment program capable of aligning a large number of genomic regions.
Our progressive alignment approach incorporates the following ideas: maximum-likelihood inference of
ancestral sequences, automatic guide-tree construction, protein based anchoring of ab-initio gene predictions,
and constraints derived from a global homology map of the sequences. We have implemented these ideas in the
MAVID program, which is able to accurately align multiple genomic regions up to megabases long. MAVID
is able to effectively align divergent sequences, as well as incomplete unfinished sequences. We
demonstrate the capabilities of the program on the benchmark CFTR region  which consists of 1.8Mb of human
sequence and 20 orthologous regions in marsupials, birds, fish, and mammals. Finally, we describe two large
MAVID alignments: an alignment of all the available HIV genomes and a multiple alignment of the entire human,
mouse and rat genomes.
\end{abstract}

\section{Introduction}

The multiple alignment problem is difficult for many reasons (see, for example, \cite{Notredame3}), and
thus remains unsolved despite much progress over the past two decades. To appreciate the complexity of the problem,
it is instructive to observe that five DNA sequences of length five have approximately
$1.05 \cdot 10^{18}$ alignments! \cite{Slowinski} Even three sequences of length five have over 14 billion alignments,
and thus it is clear that an alignment of the human, mouse and rat genomes must be based on a relatively simple
optimization criteria, and even then must involve heuristics to reduce the complexity of the problem.

Despite the overwhelming complexity of the problem, it is important to observe that
the multiple alignment problem for genomic sequences is not equivalent to the mathematical 
problem of producing an optimal alignment maximizing some score function \cite{Gusfield}. The {\em biological} problem
consists of correctly aligning homologous bases to each other, thus correctly identifying conserved non-coding
regions in introns and intergenic regions, exons in orthologous genes, and groups of orthologous genes that form
larger blocks of homology.

In this paper, we propose a method capable of rapidly aligning multiple large genomic regions
by incorporating biologically meaningful heursitics, with theoretically sound alignment strategies.
The core of our approach is a probabilistic ancestral alignment scheme \cite{Feng1, Gonnet, Holmes, Hein2, Loyt}. This involves the
progressive alignment of ancestor sequences (inferred using maximum-likelihood estimation within a probabilistic 
evolutionary model \cite{Felsenstein}) along a phylogenetic guide tree. Although a comprehensive
review of progressive alignment is beyond the scope of this paper, it is important to point
out that probabilistic approaches have been proposed and implemented (e.g. \cite{Holmes, Hein2}), although 
existing methods are not scalable to very large problems. 

 In order to incorporate biological
information into the alignment procedure, the progressive alignment is constrained by gene-based anchors.
These anchors are pre-computed based on {\em ab-initio} gene predictions and their protein alignments, and 
form part of the input to the program. In addition, non-trivial positional constraints \cite{Myers, Hardison}
are pre-computed and ensure that the progressive alignment steps respect a pre-computed homology map for the sequences. 

The alignment of the ancestor sequences is based on the AVID \cite{Bray} alignment method, thus allowing
for the rapid alignment of very large genomic sequences, even in between gene anchors that may be far apart.
This fast alignment, along with the speedup obtained by using constraints, allows for an iterative alignment
approach alternating between the progressive alignment step and phylogenetic tree construction (based on the
alignment). In fact, as we show, it is possible to start with a random initial tree and converge to the correct
guide tree thus eliminating the need for an expensive pairwise alignment step (quadratic in the number of
sequences) at the beginning of the progressive alignment \cite{thompson}.

We have combined all these ideas into a new program called MAVID, which we used to align the human,
mouse and rat genomes. We also show that MAVID is suitable for aligning very large numbers of sequences,
and is therefore practical for the alignment of multiple HIV genomes \cite{Korber} or hundreds of
mitochondrial sequences \cite{Hernnstadt}. Finally, we demonstrate the accuracy of MAVID on the benchmark
cystic fibrosis (CFTR) gene region \cite{Green} and show that it compares favorably to existing alignment methods. 

\section{Method}

Our method consists of a core progressive ancestral alignment step, which can incorporate
pre-processed constraints (see Figure 1). \\

\begin{figure}[ht]  
\begin{center}  
 \includegraphics[scale=0.7]{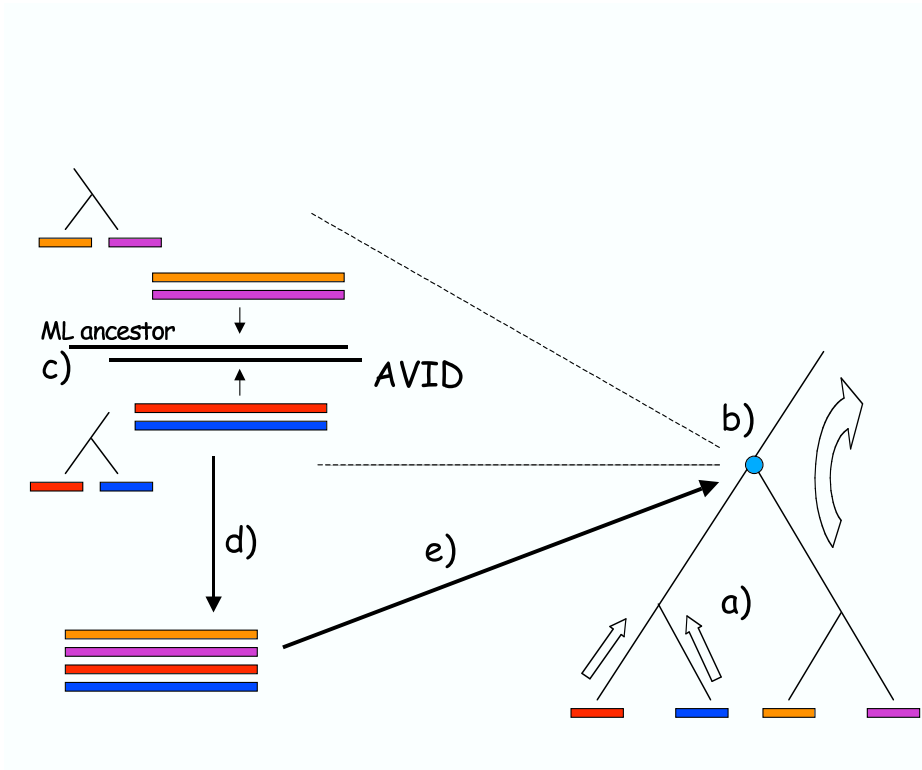}
 \end{center}  
\caption{MAVID architecture overview. (a) Sequences are aligned upwards along a guide tree and (b) alignments of alignments are performed at internal nodes. In order to align two alignments (c), maximum likelihood ancestor sequences are inferred from each of the separate alignments, and (d) the ancestor sequences are aligned with MAVID. The resulting multiple alignment (e) (corresponding to a subset of leaves of the tree) is then recorded at the internal node.}
\end{figure}

In order to clarify the presentation of the method, we begin by defining some terminology: A {\em match}
is any identified similar region between two sequences. A match does not have to be an exact at the
sequence level; for example, we could declare a match between two orthologous gene regions
even if the sequence does not match exactly. A {\em maximal exact match} is a match that is exact at the
sequence level, and is maximal (i.e. cannot be extended on either side without creating a mismatch). An {\em anchor}
is a match that is used in the alignment. A {\em constraint} $a_i \leq b_j$ in a multiple alignment means
that position $i$ in sequence $a$ must appear before position $j$ in sequence $b$ in the multiple alignment.

\subsection{Progressive Alignment}

Let $T$ be a phylogenetic tree, i.e. a binary tree with branch lengths from an
evolutionary model. In order to build progressive alignments we need a root for
$T$. Sometimes the root is known because of the availability of an outlier. In
the case in which the tree $T$ is unrooted, we use the midpoint method: select
the root to be halfway in between the two leaves of $T$ which are farthest
apart. This method produces the correct root if the molecular clock assumption
holds, and a good approximation otherwise.

We associate multiple alignments to the vertices of $T$ recursively, starting from
the leaves. For a vertex $x$ in $T$, let $T_x$ be the subtree consisting of $x$
and all the vertices beneath $x$. If $x$ is a leaf then $T_x$ is a ``trivial
tree''(i.e. a tree consisting of only one vertex) and we will label $x$ with
the sequence corresponding to $x$ in the phylogenetic tree. This sequence can
be considered a ``trivial multiple alignment''(i.e. an alignment of only one
sequence). If $x$ is an internal node then it has two children, $u$ and $v$,
which are labeled with the multiple alignments $A_u$ and $A_v$, respectively.
We then construct an alignment of all the sequences in $T_x$ by aligning the
two alignments $A_u$ and $A_v$. This procedure is applied recursively so the
program works its way from the leaves of the phylogenetic tree to the root, at
which point it will have constructed an alignment for all the sequences in $T$.

The key difference between our progressive alignment schema, and more standard
methods, is that instead of aligning $A_u$ and $A_v$ directly, we first infer
ancestral sequences $s_u$ and $s_v$ using standard phylogenetic models for inference
of the common ancestor \cite{Felsenstein}. We used the general reversible model, 
with the rate matrices obtained by \cite{Yap}.

The discussion above ignores the issue of gaps. Gaps can be modeled as a fifth
symbol which is equivalent to assigning a linear gap penalty. We have
implemented the procedure in this form, but affine gap penalties are
preferable. Furthermore, it is desirable to infer the deletion or insertion of bases in the
ancestor, and models for this already exist (e.g. TKF \cite{Thorne}). For the human-mouse-rat
alignment the issue of properly scoring gaps while inferring the  ancestral sequence was
not critical, and so we did not score with a sophisticated model, however future
work will build on probabilistic insertion/deletion models developed in \cite{Thorne0,Thorne} 
and which have already been used to develop multiple alignment algorithms \cite{Holmes3}.
It is important to note that in the MAVID alignment scheme gaps also play a role
in Smith-Waterman alignments of the ancestral sequences (see below).

After the ancestral sequence calculation, $s_u$ and $s_v$ are aligned with AVID
\cite{Bray}. AVID is a hierarchical global pairwise alignment program that
iteratively anchors exact matches and wobble matches (i.e. matches which are exact
except for possible mismatches every third base) until a final Smith-Waterman alignment step
of remaining regions. In the Smith-Waterman phase of AVID, the match and mismatch scores are
again assigned according to a substitution matrix corresponding to the branch length between
$u$ and $v$ (using the same rate matrix as for the ancestral inference). Gap scores were assigned using the 
AVID protocol, however both the gap open and gap extension scores were scaled according to the evolutionary distance.

The alignment of the ancestral sequences is then used to "glue" together $A_u$
and $A_v$ to produce a new multiple alignment which is assigned to the vertex
$x$. In particular, if position $i$ in $s_u$ matches position $j$ in $s_v$,
then column $i$ in the multiple alignment assigned to $u$ is aligned with
column $j$ in the multiple alignment assigned to $v$. Gaps in
the ancestral sequence alignments lead to gaps in the multiple alignment in the
obvious way. The procedure terminates with a final pairwise alignment at the
root node of the tree.

\subsection{Exon Anchoring and Constraints}

Our gene matches and constraints are based on a homology map for the sequences; this is a map 
that identifies the order and orientation of matching gene runs between the sequences (Colin Dewey,
manuscript in preparation). First, pairwise gene matches are computed between all the sequences.
Gene predictions are generated using GENSCAN \cite{Burge}, and every pair of predicted genes from
every pair of sequences is aligned using the translated BLAT tool \cite{Kent}. GENSCAN was selected
because it is sensitive, and BLAT provides a fast way of obtaining protein alignments between large
numbers of sequences.
 It should be noted that these programs can easily be replaced for different types
of organisms, for example a viral gene finding program is more suitable for virus alignments. 

Genes are considered to match if they form a reciprocal best hit. The genes matches are assembled into
runs which then form the basis of the homology map. Genome sequence coordinates for {\em exon} matches
are inferred from the protein alignments, thus producing a set of pairwise matches between all the sequences.
These matches are used in the obvious way: when aligning $A_u$ and $A_v$ at node $x$, all matches
that are between a sequence in $u$ and a sequence in $v$ are collected. Every such match can be converted
into a match between the ancestral sequences $s_u$ and $s_v$ which is then used in the AVID alignment.

In addition to anchoring the alignment of the ancestral sequences, the exon matches can be used in more subtle
ways to shape the final multiple alignment. It is illustrative to consider one base-pair anchors, i.e. single
matches between the sequences. Suppose we have sequences $a$, $b$, and $c$, and that $a_i$ is anchored to $c_x$
and $b_j$ is anchored to $c_y$. If we are aligning sequences $a$ and $b$ then the given anchors to $c$ do not
allow us to anchor the alignment, but they do allow us to constrain it. If $x$ is less than (resp. greater than)
$y$, we must have that $a_i$ comes before (resp. after) $b_j$ in the alignment of $a$ and $b$ if we are to produce
an alignment of $a$, $b$, and $c$ which is consistent with both of the anchors. In the language of Miller and Myers
\cite{Myers}, the two anchors provide explicit constraints on the alignment (namely that $a_i \leq c_x, a_i \geq c_x, b_j
\leq c_y,$ and $b_j \geq c_y$) but they also provide implicit constraints which are implied by transitivity:
if $x \leq y$ then we have $a_i \leq c_x \leq c_y \leq b_j$, and so $a_i \leq b_j$. This information can be
used in the alignment of the ancestral sequences by requiring potential anchors between the sequences to 
satisfy the constraints.

Thus when constructing the multiple alignment at node $x$, every triplet of sequences $(a,b,c)$ with $a$ in $u$, $b$
in $v$ and $c$ not in $x$ provides a potential constraint for the alignment. This can lead to a combinatorial explosion
of constraints. If there are $n$ sequences in the alignment then there are $O(n^3)$ such triplets, each of which
may imply many constraints. Fortunately we do not need to find the set of all possible constraints, many of which will
be redundant. Instead we wish to find a set of prime constraints (i.e. a set such that no constraint is implied by
the others) which is equivalent to, but potentially much smaller than, the set of all constraints implied by the
gene matches. Such a set can be inferred from the homology map. If there are $m$ sets of orthologous exons (not all
of which will be in every sequence) then at node $x$ there can be at most $O(m)$ prime constraints and a prime set
which is equivalent to all possible constraints can easily be found in $O(mk)$ time where $k$ is the number of
leaves below $x$. Thus all the set of all prime constraints can be found in $O(mk^2)$ time with a small constant
factor. Matches between the ancestral sequences which are inconsistent with this set of constraints can then be filtered
out in time $O(N log m)$ where $N$ is the total number of matches. For typical values of $m$ and $k$, the time taken
computing and utilizing the constraints is neglible.

Figure 2 shows an example of a constraint, and how it is enforced in the AVID alignment of the ancestral sequences.

\begin{figure}[ht]  
\begin{center}  
 \includegraphics[scale=0.7]{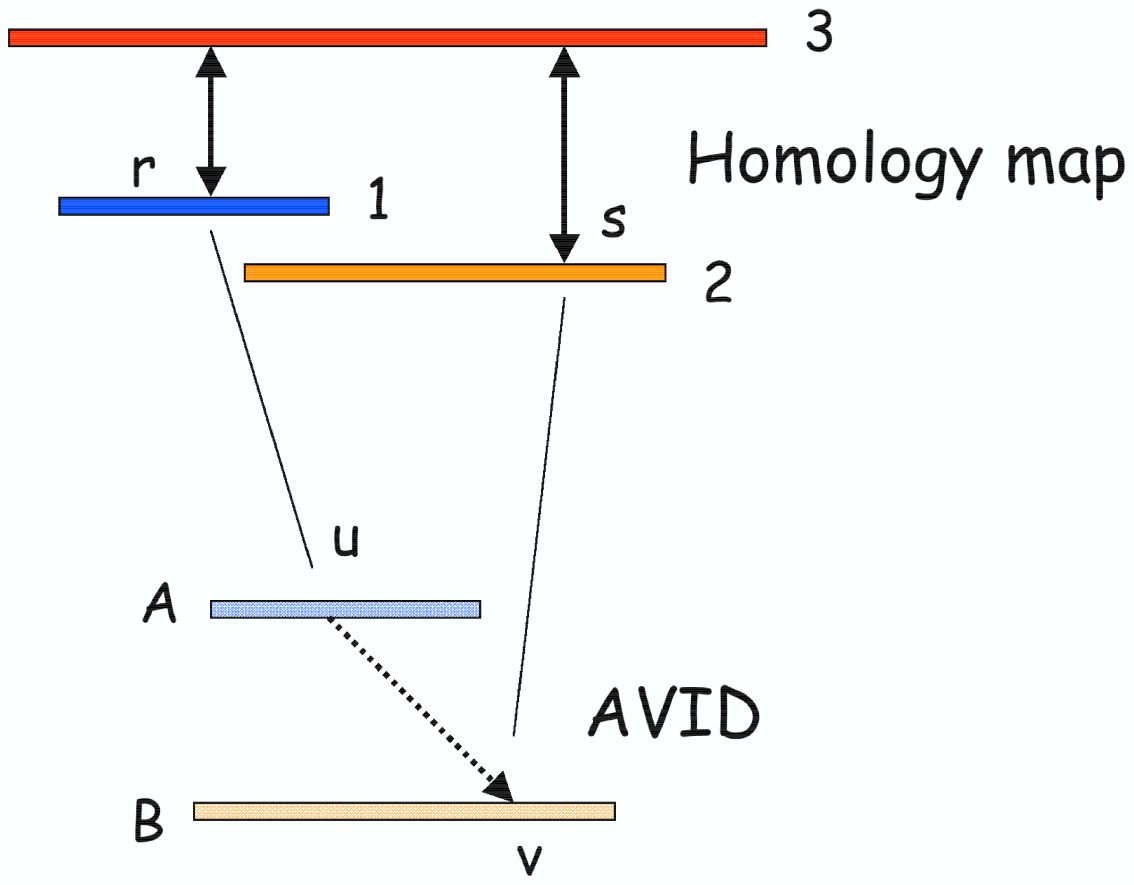}
 \end{center}  
\caption{The top half of the figure shows two exon matches determined from the homology map. In particular, exon r in sequence 1 is aligned to an exon in sequence 3, and exon s in sequence 2 is aligned to another exon in sequence 3 (double arrows). At this stage, none of the sequences have been aligned - the matches are based on the pairwise protein alignments of the predicted genes. During an MAVID alignment of ancestral sequences in the progressive multiple alignment, position r from sequence 1 maps to position u in the ancestral sequence A, and position s maps to position v in the ancestral sequence B (solid lines). Even though sequence 3 is not in the multiple alignment yet, the constraint forces position u to be aligned before position v in the final multiple alignment (dashed line). The constraint is enforced by removing all the matches violating the constraint from consideration during the anchoring of the alignment. 
}
\end{figure}

The pre-processing step of finding all exon matches is quadratic in the number of sequences, however since the
protein alignments are gene based, they are typically computed on less than $5\%$ of the sequence. Thus, the
gene matching is actually significantly faster than translated match finding which requires searching the
entire sequence in all three frames and on both strands. Furthermore, by comparing only the proteins produced
by a gene prediction program, the program implicitly takes into account splice sites and other gene features
in building gene anchors. It is also important to note that this approach is completely {\em ab initio} even though
a gene finding step is necessary, no information beyond the sequences is used. For this paper we performed
alignments using this strategy in order to demonstrate the performance of an {\em ab initio} approach. However it
is possible to make use of mRNA and EST data, thus incorporating known biological annotation about the sequences
into the alignment.

\subsection{Tree Building}

Most multiple alignment programs require pairwise alignments of all the sequences to build an initial guide
tree. This step requires a quadratic number of sequence alignments and is infeasible for large numbers of
sequences. We utilize an iterative method to obtain a guide tree using only a linear number of alignments.

The initial guide tree is selected randomly from the set of complete binary trees (or almost complete
binary trees in the case where the number of sequences is not a power of 2). For a given number of
nodes, these are the binary trees with minimal depth and thus initial errors in pairwise alignments
have less opportunity to propagate through the tree. The sequences are aligned using this random tree and
then a phylogenetic tree is inferred from the resulting multiple alignment. The likelihood of the tree given
the alignment can be used as a quantitative measure of the quality of the tree and the process is iterated
until the alignment and tree are satisfactory.

For small numbers of sequences, the inference of the tree from the multiple
alignment can be done using maximum-likelihood methods and accounts for only a
small percentage of the running time. However as the number of sequences
increases, we have found that ML reconstruction becomes impractical and
neighbor-joining must be used. Because pairwise alignments are easy to infer
from a multiple alignment, we can perform neighbor-joining reconstruction
rapidly even with large numbers of sequences. We have tested MAVID with the
fastDNAml \cite{Olsen} program for smaller datasets and the CLUSTALW
implementation of neighbor joining for larger problems.

Instead of computing all pairwise alignments, only $O(nk)$ alignments are
necessary to perform $n$ iterations with $k$ sequences. We found that for typical 
alignment problems only a small number of iterations were necessary (see results
on the HIV sequences below). 
It is important to note that our iterative method (multiple alignment alternating with neighbor joining) 
is considerably less sophisticated than ML methods such as SEMPHY \cite{Friedman} or MCMC sampling methods that
search through combined alignment/tree space \cite{Hein}. However our 
approach is scalable to large problems and, as we have pointed out, appears to converge quickly in practice.

\section{Results}

\subsection{A human, mouse and rat whole genome multiple alignment}

We aligned the human (April 2003), mouse (February 2003), and rat (June 2003) genomes
using MAVID. A homology map for the genomes was built by Colin Dewey (manuscript in preparation), and
was used to generate gene anchors and constraints. Figure 3 summarizes the exon coverage of the
alignment on chromosome 20: it shows how many of the RefSeq genes were covered by anchors (and
therefore automatically aligned correctly), and how many were subsequently aligned by MAVID. Chromosome
20 was chosen because it aligns almost completely with mouse chromosome 2, and therefore the quoted
numbers should be useful for comparing MAVID to other alignment approaches which do not
explicitly separate out orthologous from paralogous alignments.
\begin{figure}[ht]  
\begin{center}  
 \includegraphics[scale=0.7]{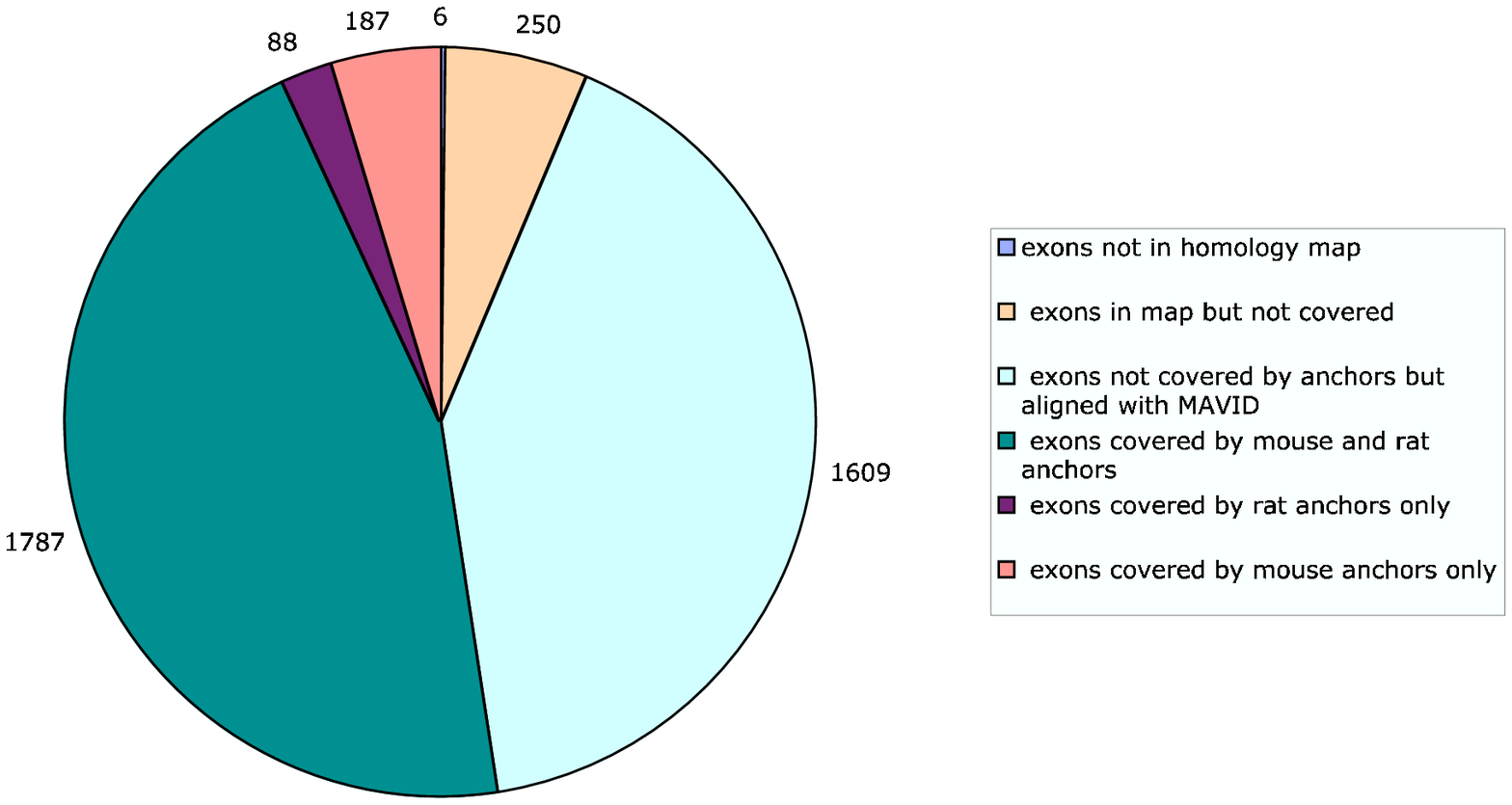}
 \end{center}  
\caption{Coverage of human chromosome 20 RefSeq exons by the MAVID alignments. Out of a total of 3927 exons, only 6 were not in the homology map. 53.5\% of the exons were covered by pre-computed exon anchors in either mouse or rat. The remaining exons are mostly aligned by MAVID, resulting in 93.6\% of the exons covered by alignment in either mouse or rat.}
\end{figure}

The MAVID alignments have been used to estimate evolutionary rates for the genomes, and to identify
evolutionary hotspots where one of the rodent genomes has been evolving much more slowly than the
other (these results are reported in another companion paper \cite{Yap}).  They 
are also used to support the K-BROWSER \cite{Kushal}, which is a new browser especially
designed to view multiple genomes, their associated annotations and alignments. 

\subsection{CFTR region: 21 organisms}
We aligned ~1.8Mb of human sequence together with the homologous regions from
20 other organisms (baboon, cat, chicken, chimp, cow, dog, dunnart, fugu, hedgehog, horse, lemur,
macaque, mouse, opposum, pig, platypus, rabbit, rat, tetraodon and zebrafish) for a total of
23Mb of sequence. This sequence has been generated by NISC as part of a comprehensive project to
sequence a number of regions in the genome in multiple organisms for evolutionary and functional
studies. However it is important to note that some of the sequences remain incomplete, contributing
to the difficulty of the multiple alignment problem. A subset of this data-set (13 organisms)
has recently been used \cite{Brudno} as a benchmark to compare pairwise and multiple alignment
programs.

The map-building step takes approximately 15 minutes on a 2.6GhZ processor, with peak memory usage
of roughly 700Mb for GENSCAN. The subsequent MAVID alignment takes another 24 minutes, for a
total of about 40 minutes. The tree reconstruction step takes less than a minute using neighbor
joining. Thus, an iterative approach to building the tree is feasible, and a stable tree was
constructed after only two rounds of alignment.

It is difficult to assess the overall quality of the alignment, but one feature which can be
verified is the alignment of exons. In order to do so, we projected the alignment onto the human
sequence in order to produce pairwise alignments between human and each of the other 20 sequences.
This analysis was complicated by the fact that the sequencing is not complete and so not every
exon has been sequenced in every organism. In order to address this shortcoming, we 
calculated the fraction of human exons which were aligned with each of the sequences. An exon was
considered to be aligned if at least 70\% of it was covered by alignment, and at least 50\% of
the bases were matching.

The MAVID alignments were compared to MLAGAN, version 1.1 \cite{Brudno}. MLAGAN is the only other program we
know of that is able to align the 21 sequences in a reasonable period of time (the running time of MLAGAN
on the 21 sequences is roughly 6 hours). DIALIGN \cite{morgenstern2}, also designed for large
genomic regions, was too slow for processing the sequences: even with the new CHAOS/DIALIGN program,
aligning only 4 of the sequences took 14 hours. The MGA program \cite{Hohl} is designed for very similar
sequences only, and so was not suitable for the diverse fish, bird, marsupial mammal alignments.\\
\begin{center}  
 \includegraphics[scale=0.7]{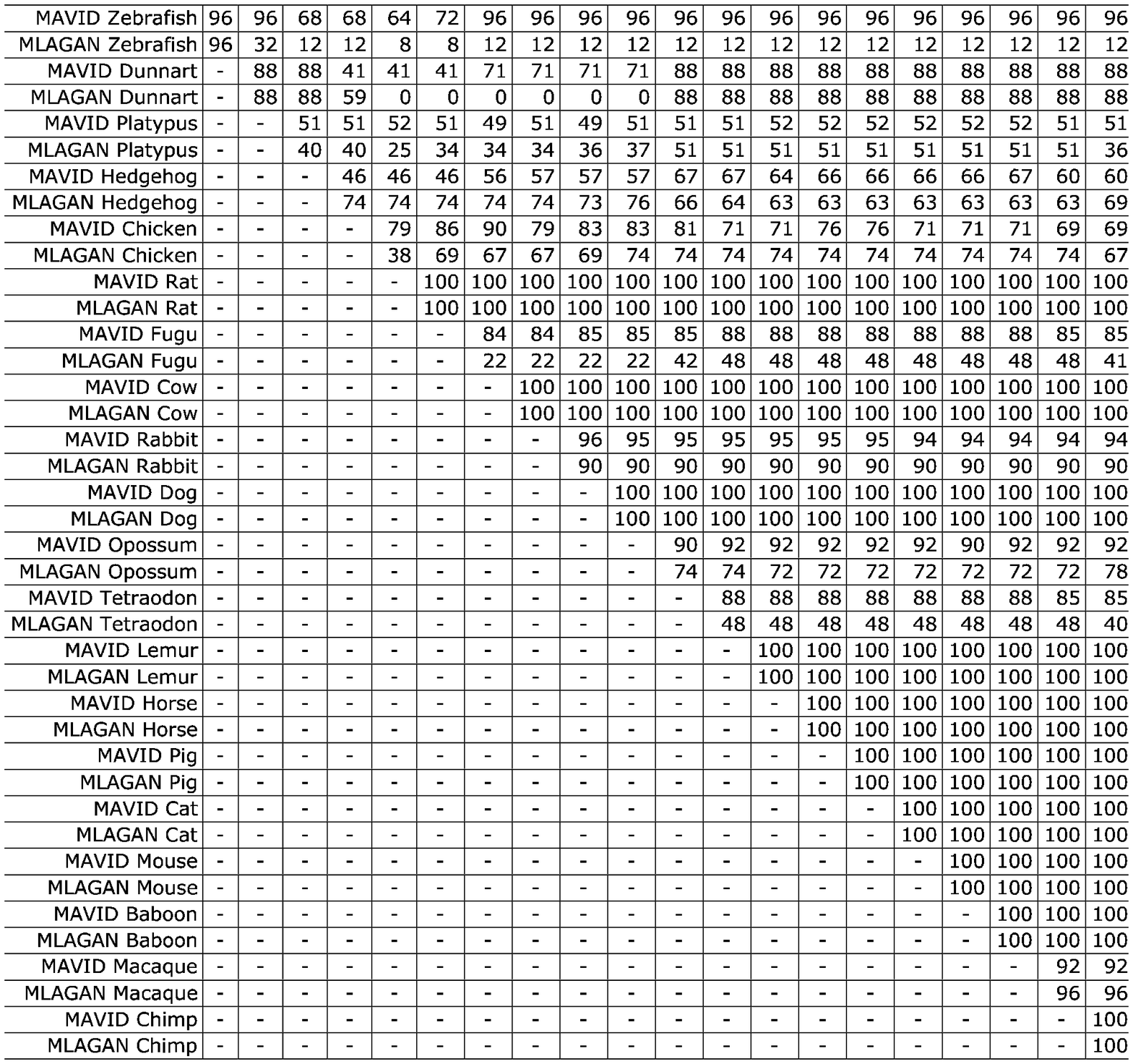}
 \end{center}  
Table1: Comparison of MLAGAN and MAVID multiple alignments on the 21 organism CFTR alignment. Each row in the table shows the \% coverage of the alignable exons in an organism as calculated by extracting the pairwise alignment with human from the multiple alignment. The different columns in the table correspond to the number of sequences in the multiple alignnment. Thus, the first column corresponds to the pairwise alignment of human-zebrafish and the last column to the multiple alignment of all 21 sequences).  Organisms were added according to the k-MST, so that the alignment problems are as difficult as possible (mutually most distant organisms).

In order to better understand how alignment accuracy varies with the number of sequences
being aligned, we compared the MAVID and MLAGAN alignments on 21 different {\em datasets}, beginning 
with a pairwise alignment of the most distant organisms, and adding in one (mutually most distant) organism
at a time all the way up to a comparison on the 21 sequences. In order to do this we computed the
human clamped k-MST trees \cite{Boffelli}, i.e. the subtrees on $k$ leaves
of maximum weight with the human sequence as one of the leaves. Thus, alignments were computed first
for human and zebrafish alone, then for human, zebrafish and dunnart, and eventually all of the 21 sequences.
The order in which they were added was: human, zebrafish, dunnart, platypus, hedgehog, chicken, rat, fugu,
cow, rabbit, dog, oppossum, tetraodon, lemur, horse, pig, cat, mouse, baboon, macaque, chimp. Exon coverage 
was calculated by first running TBLASTX to identify human exon homologs in the other species using the same
criteria as in \cite{Brudno}, and then computing coverage with respect to the identified exon sets.

The results of the alignments show that both programs correctly aligned mammalian sequences. The
alignment of distant organisms shows much greater variability with respect to the sequences included in
the alignment problem. For example, adding fugu to an alignment of human, zebrafish, and dunnart may improve
the alignment, but as Table 1 demonstrates, adding platypus can degrade it. MAVID shows significant improvement
over MLAGAN in this respect.

\subsection{HIV1/SIV: complete genomes from 242 individuals}

The HIV databases maintained at LANL contains a collection of HIV-1, HIV-2, and
SIV sequences, carefully linked with individuals and their histories. We
extracted the complete genomic sequences of HIV1 and SIV from this database
(currently totaling 242) and aligned them with MAVID. The alignment of the
sequences takes 2.5 minutes. A phylogenetic tree was constructed with neighbor
joining taking an additional 30 seconds. Again, it is difficult to assess the quality
of the alignment, but it is accurate enough that the different strains cluster
in the inferred tree (see Figure 4). In order to understand the stability of 
the tree building/alignment iteration, we examined 100 alignment runs 
on a reduced sequence set with the recombinant strains removed (since recombination is not addressed by standard phylogenetic 
models).
We found that the correct strains were 
grouped together within one round of alignment (starting with a random tree) in all the 100 runs.
To validate the iterative alignment/tree building procedure, we also examined 
the number of sub-trees that were fixed with each successive round of alignment. 
The tree inferred after the  second round of alignment has $45.87$ of its 
subtrees in agreement with the tree after the first run (on average), and 
the number fixed between the second and third rounds is $54.08$ on average.
Starting with different random trees and aligning for three rounds, we find
that $54.16$ of the sub-trees agree on average. Our criteria for comparing trees (exact agreement of sub-trees)
is rather strict, and these numbers are very encouraging (one would expect about one non-trivial sub-tree
to agree for two random trees). 
Although a MAVID multiple alignment
combined with a neighbor joining tree may not be as accurate as hand edited
alignments followed by maximum likelihood tree building, it can serve to
provide very fast results that can then form the basis for further
refinement. An alignment problem of this size is not practical on a standard 
desktop computer
if all of the pairwise alignments are computed in order to build an initial guide tree (as is done 
in many programs, e.g. CLUSTALW).

\begin{figure}[ht]  
\begin{center}  
 \includegraphics[scale=0.7]{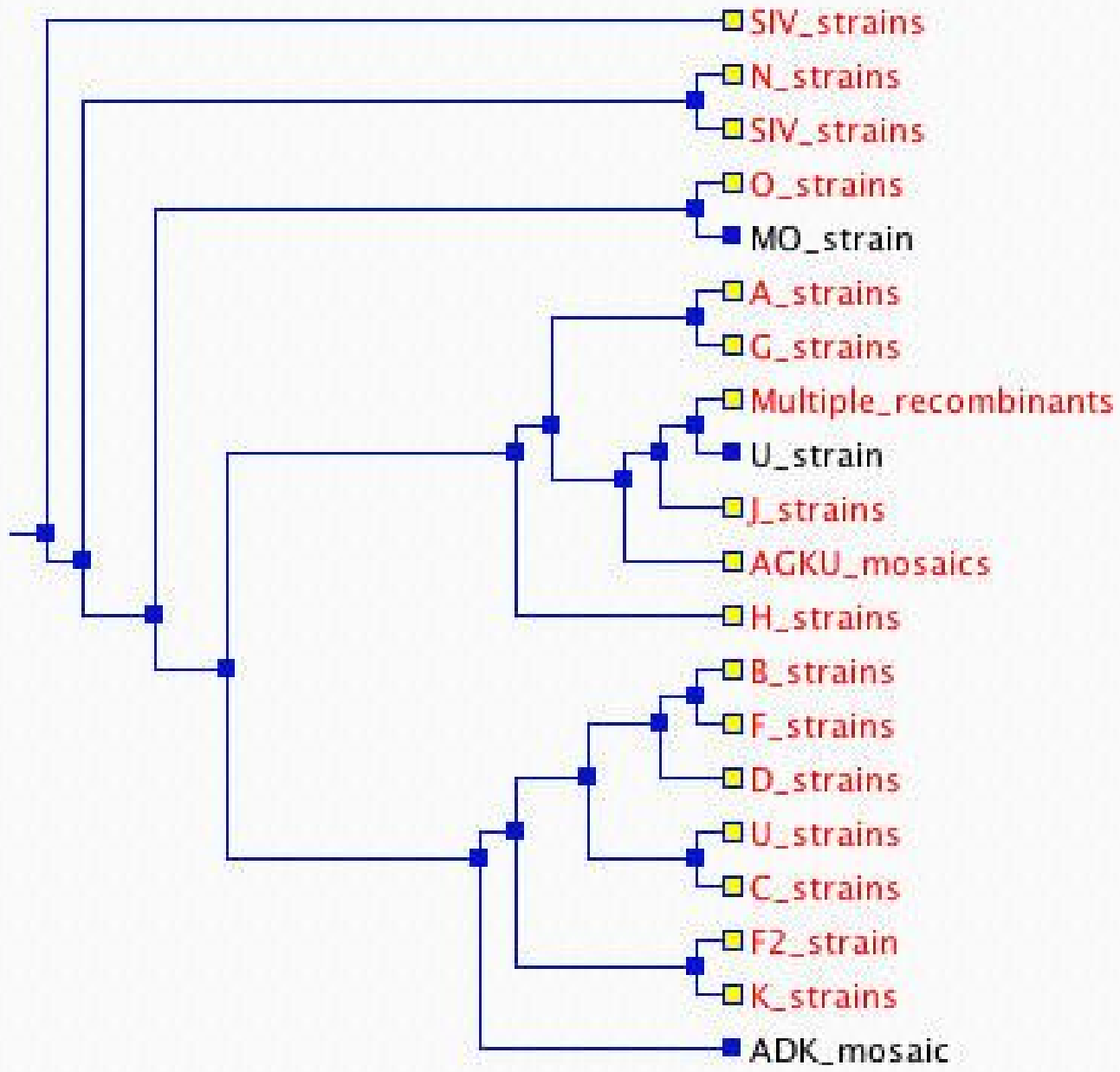}
 \end{center}  
\caption{The HIV tree as inferred from 242 sequences obtained from Los Alamos National Laboratories. The sequences are labeled by strains, and the different strains have been grouped together (yellow leaves). }
\end{figure}

The alignments and phylogenetic trees for all the above sequences are
downloadable at \\
{\tt http://baboon.math.berkeley.edu/mavid/data}

\section{Conclusion}

As we have outlined in the introduction, we view the genomic multiple sequence alignment problem as a 
{\em biological} alignment problem, rather than a purely mathematical one. That is, the incorporation
of biologically relevant information (in our case ab-initio gene predictions) is critical to building accurate
alignments and correctly identifying homologous relationships. Our method of incorporating constraint
information into the alignments helps address one of the primary objections to progressive alignment
strategies, namely that progressive alignment is ``local'' with the alignment at each node containing
only information about the sequences below it. The application of constraint information can be thought
of as a "look ahead" step which helps to fix potential problems.

Our approach is also consistent with a number of other ideas which we have not yet implemented, but
which could be easily integrated into MAVID and will improve results. Iterative refinement, the process
of re-aligning across an edge in the tree, fits in naturally with our framework  \cite{Gotoh1,Gotoh2}.
Similarly, the homology map which MAVID uses can indicate information about inversions and duplications,
and this can be used to correctly align regions containing rearrangements.

MAVID compares favorably with existing programs. As we have pointed out, it is significantly more accurate
than MLAGAN on the alignment of the CFTR benchmark region. MLAGAN is the only other program we know of
that can even align such a large dataset. We also know of no other programs that can quickly align hundreds of
viral or mitochondrial genomes.

A MAVID web server has been operational for over six months and processes over 700 requests a month \cite{Bray2}.
Alignment requests have ranged from large genomic regions in mammals, fish, flies and plants, to alignments of
viruses, mitochondria, and other bacterial genomes. MAVID can be accessed at
{\tt http://baboon.math.berkeley.edu/mavid/}. The program is freely available for academic and non-profit use. 

\section{Acknowledgments}

We thank Colin Dewey for providing access to his homology map making programs. We thank Von Bing Yap for
helping with the evolutionary models used in MAVID. Thanks to Ingileif Brynd\'{\i}s Hallgr\'{\i}msd\'ottir
for her help throughout the project and for her comments on the final manuscript. The data used in the
multiple alignment of the CFTR region was generated by the NIH Intramural Sequencing Center (www.nisc.nih.gov),
and was used subject to their six month hold policy. The HIV sequences were downloaded from the HIV database
(hiv-web.lanl.gov). Thanks also to the rat sequencing consortium, both for providing the rat sequence to
align, and for facilitating helpful collaborations and discussions. Finally, we thank the anonymous reviewers for
their insightful comments and suggestions. This work was partially supported by funding from the NIH (grant R01-HG02362-01) and the Berkeley PGA grant from the NHLBI.

\section*{Website References}

http://www.nisc.nih.gov/ -- NIH Intramural Sequencing Center\\
http://hiv-web.lanl.gov/ -- LANL HIV Databases\\
http://baboon.math.berkeley.edu/mavid/ -- The MAVID Webserver\\
http://baboon.math.berkeley.edu/mavid/data/ --  Supplemental Data\\
http://hanuman.math.berkeley.edu/kbrowser/ -- KBROWSER\\

\end{document}